\documentclass[preprint,superscriptaddress,showpacs,nofootinbib,floatfix,natbib]{revtex4-1}
\usepackage[utf8]{inputenc}
\usepackage[dvipdfmx]{graphicx}
\usepackage[usenames]{color}
\usepackage{setspace}
\usepackage{amsmath}
\usepackage{amssymb}
\usepackage{amsthm}
\usepackage{bm}
\usepackage{hyperref}
\usepackage{slashed}

\begin{document}

\preprint{UT-HET-134}

\title{Higgs potential in gauge-Higgs unification with a flat extra dimension}

\author{Mitsuru Kakizaki}
\email{kakizaki@sci.u-toyama.ac.jp}
\affiliation{
  Department of Physics, University of Toyama,
  3190 Gofuku, Toyama 930-8555, Japan
}
\author{Shin Suzuki}
\email{suzuki@jodo.sci.u-toyama.ac.jp}
\affiliation{
  Department of Physics, University of Toyama,
  3190 Gofuku, Toyama 930-8555, Japan
}

\date{July 28, 2021}

\begin{abstract}

We analyze the structure of the Higgs potential in gauge-Higgs unification 
with a flat extra dimension.
As a concrete model, we first consider the cases where the
Standard Model Higgs doublet is embedded into a higher-dimensional $SU(3)_{w}^{}$ gauge multiplet with 
five-dimensional Lorentz symmetry relaxed.
In this $SU(3)_{w}^{}$ model with Lorentz violation, the deviation of the 
resulting triple Higgs boson coupling from its SM prediction is shown
to be less than $10\%$ when the compactification scale is larger than the
experimental lower bound, which is around 5~TeV. 
Next, we examine the Higgs potential in other similar gauge-Higgs
unification models with a flat extra dimension.
It is pointed out that even in such models
the shape of the Higgs potential around the vacuum 
quickly approaches that of the minimal Higgs potential with one Higgs doublet as the
compactification scale of the extra dimension increases.
An observable deviation of the triple Higgs boson coupling at planned lepton colliders
will necessitate significant extensions of such gauge-Higgs unification models.

\end{abstract}

\maketitle

\section{Introduction}

The discovery of the Higgs boson $h$ with a mass of 
$m_h^{} = 125~\mathrm{GeV}$~\cite{Aad:2012tfa,Chatrchyan:2012ufa}
and the measurements of its couplings~\cite{Khachatryan:2016vau,Aaboud:2017vzb,Sirunyan:2018koj}
at the CERN Large Hadron Collider (LHC) have established  
the mechanism of the spontaneous electroweak symmetry breaking, 
on which the Standard Model (SM) is based.
Nevertheless, there still remain unknowns regarding the Higgs sector.
These include the guiding principle, the shape of the Higgs potential,
dynamics behind the electroweak symmetry breaking (EWSB), etc.
Besides, new physics beyond the SM is necessary for accounting 
for neutrino oscillations, 
baryon asymmetry of the Universe, the existence of dark matter
and cosmic inflation.
As listed above, the SM has both theoretical and phenomenological puzzles, and 
many ideas about these issues have been proposed.

One of the notorious problems the SM bears is the so-called hierarchy problem.
Since the Higgs doublet is an elementary scalar field in the SM,
quantum corrections to the mass squared of the Higgs doublet is quadratically divergent.
In order to obtain the observed mass of the Higgs boson, 
an unnaturally huge cancellation 
between the bare mass squared and the quantum corrections is required.
This fine-tuning problem can be evaded if one invokes yet-to-be-discovered
paradigms, such as supersymmetry (SUSY)~\cite{Wess:1974tw,Salam:1974yz}, compositeness~\cite{Kaplan:1983fs} and
gauge-Higgs unification (GHU)~\cite{Manton:1979kb,Fairlie:1979zy,Fairlie:1979at,Hosotani:1983xw,Hosotani:1983vn,Hosotani:1988bm}.
In SUSY models, the quadratic divergence from the SM particles are canceled out
by that from the corresponding superpartners.
In composite models, the SM Higgs boson is composed of a pair of fermions, 
and thus chiral symmetry can avoid the problem arising from an elementary scalar.
In GHU models, the SM Higgs doublet is embedded into 
some extra-dimensional component of a vector multiplet.
Due to the higher-dimensional gauge symmetry, 
the mass squared of the Higgs boson is protected against quadratic divergence~\cite{Hatanaka:1998yp,Haba:2008dr}.
In more detail, the Higgs potential is flat at the tree level due to the higher-dimensional gauge symmetry and induced by the quantum corrections at the loop level.
Although the higher-dimensional gauge theory is nonrenormalizable, the finiteness of the Higgs potential has been validated at the two loop level in some GHU models~\cite{Maru:2006wa,Hosotani:2007kn}.
The Higgs potential at higher-loop levels has also been investigated~\cite{Hisano:2019cxm}. 

Although the concept of GHU is elegant, it is difficult to obtain the measured masses
of the top quark and Higgs boson.
Challenges in solving this problem lead to a rich phenomenology 
in GHU~\cite{Scrucca:2003ra,Cacciapaglia:2005da}.
In this paper, we focus on the Higgs potential in GHU models and investigate the triple Higgs boson coupling.
The triple Higgs boson coupling in the framework of GHU have been discussed 
in a flat five-dimensional (5D) orbifold model with large bulk fermion representations~\cite{Adachi:2018gud},
and in warped models~\cite{Funatsu:2020znj}.
In contrast to these earlier works, we employ a flat orbifold model with 5D Lorentz symmetry relaxed in order to obtain realistic values of the masses
of the top quark and Higgs boson.
As an example of broken Lorentz symmetry in the fifth dimension, 
we consider the $SU(3)_{w}^{}$ model studied
in Refs.~\cite{Panico:2005dh}, compute the triple Higgs boson coupling and 
compare the predicted value with the expected precision at future lepton colliders, such as 
the International Linear Collider (ILC)~\cite{Fujii:2019zll} and the Compact LInear Collider (CLIC)~\cite{Roloff:2019crr}.
We show that the characteristics of the effective potential and triple Higgs boson coupling are
applicable to a wide class of GHU models with a flat orbifold.

This paper is organized as follows. 
In Sec.~\ref{sec:Model}, we briefly review the $SU(3)_{w}^{}$ GHU model 
on a  flat $S^1/Z_2^{}$ orbifold with 5D Lorentz symmetry relaxed,
and describe the matter Lagrangian and mass spectrum.
The calculation of the resulting effective Higgs potential 
is demonstrated in Sec.~\ref{sec:Veff}.
Phenomenological constraints on this model are given in Sec.~\ref{sec:constraints}.
Then, we show the prediction of the triple Higgs boson coupling in Sec.~\ref{sec:hhh}.
Section~\ref{sec:summary} is devoted to a summary.

\section{Model}
\label{sec:Model}

Here, we briefly outline the 5D GHU model proposed in Ref.~\cite{Panico:2005dh}.
In this model, the extra dimension is compactified on a flat $S^1/Z_2$ orbifold
with the $S^1$ compactification radius $R$, and 
Lorentz symmetry in the fifth dimension is assumed to be broken in order to obtain
realistic masses of the top quark and Higgs boson, as discussed later.
The electroweak gauge group is extended to $SU(3)_w^{} \times U(1)^{\prime}$.
The 5D $SU(3)_w$ and $U(1)^{\prime}$ gauge coupling constants, $g_5^{}$ and $g_5^{\prime}$, 
are related to the corresponding four-dimensional (4D) gauge coupling constants, $g$ and $g^{\prime}$, through
\begin{align}
g_5 \equiv g\sqrt{2\pi R}, \quad g_5^{\prime} \equiv g^{\prime}\sqrt{2\pi R}.
\end{align}
The extended electroweak gauge group $SU(3)_w \times U(1)^{\prime}$ is
broken down to $SU(2)_L \times U(1)_w \times U(1)^{\prime}$ by
the nontrivial $Z_2$ orbifold boundary condition whose projection matrix is expressed as
\begin{align}
P=
\begin{pmatrix}
	-1 & 0 & 0 \\
	0 & -1 & 0 \\
	0 & 0 & +1 \\
\end{pmatrix}  ,
\end{align}
in the $SU(3)_w^{}$ space.
After the orbifold compactification, there appear massless 4D
bosonic fields from
the $SU(2)_L \times U(1)_w^{}$ components of the $SU(3)_w^{}$ gauge fields $A_\mu^a$,
$U(1)^{\prime}$ gauge field $A_{\mu}^{\prime}$,
and doublet scalar component $A_5^a$, which is charged under $U(1)^{\prime}$ 
and plays the role of the Higgs doublet. 
The vacuum expectation value (VEV) of $A_5^a$ induces an additional spontaneous symmetry breaking to $U(1)_{\rm EM}^{}$.
It is convenient to parametrize the VEV of $A_5^a$ using a parameter $\alpha$ as
\begin{align}
\langle A_5^a \rangle =\frac{2\alpha}{g_5^{} R}\delta^{a7}  ,
\end{align}
where $\delta^{ab}$ is the Kronecker delta in the $SU(3)_w^{}$ adjoint representation.
The EWSB in this model is detailed in Sec.~\ref{sec:Veff}.

Notice that the extra $U(1)^{\prime}$ gauge group and its gauge coupling constant $g^{\prime}$ are 
necessary for modifying the Weinberg angle $\theta_W^{}$.
If the $SU(2)_L^{} \times U(1)_Y^{}$ gauge group is a subgroup of
the $SU(3)_w^{}$, an unacceptably large Weinberg angle of $\theta_W^{}=\pi/3$ is predicted.
Thanks to the existence of the $U(1)^{\prime}$ gauge group,
one can set the hypercharge generator $Y$ to the linear combination 
of the $U(1)_w$ and $U(1)^{\prime}$ generators as $Y=t_8/\sqrt{3}+t'$.
Then, the hypercharge gauge field $A_Y$ and its orthogonal gauge field $A_X$ are given by
\begin{align}
A_Y=\frac{g^{\prime}A^8+ \sqrt{3}g A^{\prime}}{\sqrt{3g^2+g^{\prime 2}}}, \quad
A_X=\frac{ \sqrt{3}g A^8+ g^{\prime}A^{\prime}}{\sqrt{3g^2+g^{\prime 2}}}.
\end{align}
The $U(1)_Y^{}$ gauge coupling constant $g_Y$ is given by a combination of the $SU(3)_w^{}$ and $U(1)^{\prime}$ coupling constants as $g_Y=\sqrt{3}gg^{\prime}/\sqrt{3g^2+g^{\prime 2}}$.
The Weinberg angle is modified to
\begin{align}
\sin^2\theta_W^{}=\frac{g_Y^2}{g^2+g_Y^2}=\frac{3}{4+3g^2/g^{\prime 2}}\,.
\end{align}
By adjusting the value of $g^{\prime}$, we can obtain the measured Weinberg angle.

\subsection{Matter Lagrangian}

Let us turn to the matter sector of this $SU(3)_w^{}$ GHU model,
where both massive bulk fermions and massless brane-localized fermions are introduced.
By adjusting mixing terms between the bulk and brane fermions,
one can reproduce the observed hierarchy of the fermion masses.
Each brane fermion can be localized at either of the two fixed points,
$y=0$ or $y=\pi R$ of the $S^1/Z_2^{}$ orbifold.
In our analysis, we locate the left-handed and right-handed brane fermions at the separate fixed
points 
in order to reconcile the masses of the top quark and Higgs boson with the observed values.
This choice corresponds to the $\delta=1$ set-up in Ref.~\cite{Panico:2005dh}.
In our analysis, we focus on the third generation quarks
because radiative contributions to the Higgs potential from the first and second generation quarks as well as those from leptons are relatively small.
We introduce bulk fermions that are coupled to the brane top (bottom) quark.
These bulk fermions are assigned to
a symmetric (fundamental) representation of $SU(3)_w^{}$
and periodic on $S^1$.
In addition, in order to adjust the shape of the Higgs potential,
one needs another set of bulk fermions that are symmetric representations 
of $SU(3)_w^{}$ and antiperiodic on $S^1$.

\begin{figure}[t]
 	\begin{center}
		\includegraphics[clip,width=12cm]{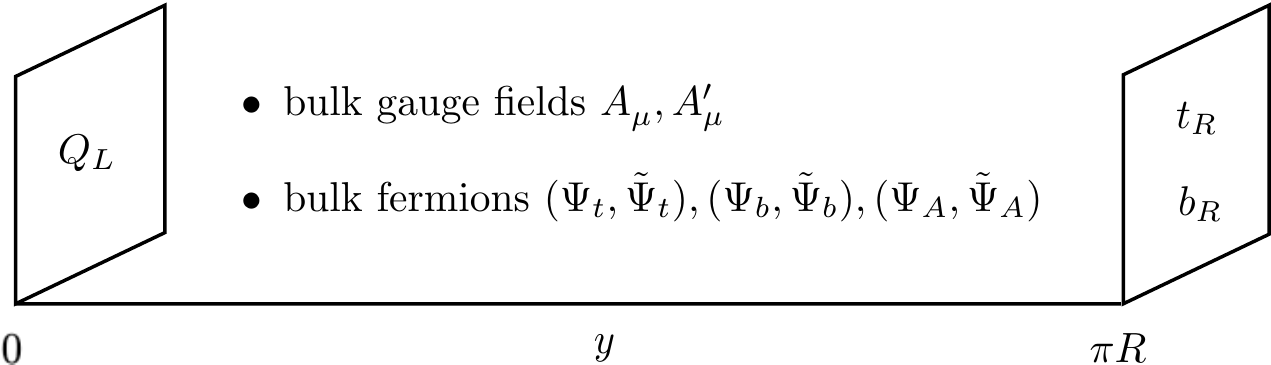}
		\caption{Field configuration on the $S^1/Z_2^{}$ orbifold.}
		\label{fig:layout}
	\end{center}
\end{figure}

Now we are in a position to specify the configuration of the above-mentioned bulk and
brane-localized fermion fields.
There are three types of a pair of colored 5D bulk fermions as follows:
periodic bulk fermions $(\Psi_t^{},\tilde{\Psi}_t^{})$
in the $(\bm{3},\bar{\bm{6}})$ representation;
periodic bulk fermions $(\Psi_b,\tilde{\Psi}_b)$ in the $(\bm{3},\bm{3})$ representation; and
antiperiodic bulk fermions $(\Psi_A^{},\tilde{\Psi}_A^{})$ in the $(\bm{1},\bm{6})$ representation.
Here, the numbers in the parentheses denote representations under $SU(3)_c \times SU(3)_w$.
The hypercharge of $(\Psi_A^{},\tilde{\Psi}_A^{})$ 
is chosen such that $(\Psi_A^{},\tilde{\Psi}_A^{})$ do not mix with the bulk or brane leptons,
which are necessary for the lepton masses.
For each pair of the bulk fermions, opposite $Z_2$ parities are assigned 
so that the mixing terms between $\Psi_j$ and $\tilde{\Psi}_j$ ($j=t,b,A$)
with 5D mass parameters $M_j$ are allowed.
One of the characteristic features of this model is that
$SO(4,1)$ Lorentz invariance is relaxed.
Parameters $k_j$ and $\tilde{k}_j$ are introduced
to parametrize such effects residing in the kinetic terms for bulk fermions $\Psi_j^{}$ and $\tilde{\Psi}_j^{}$, respectively.
A left-handed $SU(2)_L$ doublet $Q_L=(t_L,b_L)^{\rm T}$ 
are localized at the $y=0$ brane, and right-handed $SU(2)_L$
singlets $t_R$ and $b_R$ at the $y=\pi R$ brane, respectively.
The matter contents are summarized in Table~\ref{tab:matter}.
The field configuration on the $S^1/Z_2{}$ orbifold is depicted in Fig.~\ref{fig:layout}.
We introduce parameters $e_{1}$ and $e_{2}$ with mass dimension half in order to parametrize
boundary couplings for mixing terms between the bulk and brane fermions.
From the remarks above, the 5D matter Lagrangian is written as
\begin{align}
\mathcal{L}_{\rm matter}=&\sum_{j=t,b,A}\left\{
\bar{\Psi}_j\left(i\slashed{D}_4-k_jD_5\gamma^5\right) \Psi_j
+\bar{\tilde{\Psi}}_j\left(i\slashed{D}_4-\tilde{k}_jD_5\gamma^5\right)\tilde{\Psi}_j
+\left(\bar{\Psi}_j M_j \tilde{\Psi}_j+{\rm h.c.} \right) \right\}
\nonumber
\\
&+\delta(y-0)\left\{\bar{Q}_L i\slashed{D}_4Q_L
+\left(e^b_1\bar{Q}_L\psi_b +e^t_1\bar{Q}^c_R\psi_t+{\rm h.c.} \right) \right\}
\nonumber
\\
&+\delta(y-\pi R)\left\{\bar{t}_R i\slashed{D}_4 t_R +\bar{b}_R i\slashed{D}_4 b_R
+\left(e^b_2\bar{b}_R \chi_b +e^t_2 \bar{t}^c_L \chi_t +{\rm h.c.} \right) \right\} ,
\end{align}
where $\slashed{D}_4 \equiv \gamma^{\mu}D_{\mu}$ with 
$D_\mu$ and $D_5$ being covariant derivatives, and
$\psi_{t,b}$ and $\chi_{t,b}^{}$ are the $SU(2)_L^{}$ doublet and singlet components of the bulk fermions $\Psi_{t,b}^{}$.
The SM fermions are approximately given by the brane fermion components.

\begin{table}[t]
\caption{Matter contents and the quantum numbers. The hypercharge of $(\Psi_A,\tilde{\Psi}_A)$
is chosen such that $(\Psi_A,\tilde{\Psi}_A)$ do not mix with the bulk or brane leptons. The color factor is denoted by $C_F^{}$.}
\label{tab:matter}
\centering
\begin{tabular}{|c||c|c|l|c|} \hline
	Fields & $SU(3)_c \times SU(3)_w$ &  periodicity ($\eta$) \  & $SU(3)_c \times SU(2)_L \times U(1)_Y$ &\  $C_F$ \  \\ \hline \hline
	$(\Psi_t, \tilde{\Psi}_t )$ & $({\bf 3},{\bf \bar{6}})$ & periodic (0) & $({\bf 3},{\bf 1})_{2/3} +({\bf 3},{\bf 2})_{1/6} +({\bf 3},{\bf 3})_{-1/3}$ & $3$ \\ \hline
	$(\Psi_b, \tilde{\Psi}_b )$ & $({\bf 3},{\bf 3})$ & periodic (0) & $({\bf 3},{\bf 1})_{-1/3} +({\bf 3},{\bf 2})_{1/6}$ & $3$ \\ \hline
	$(\Psi_A, \tilde{\Psi}_A )$ & $({\bf 1},{\bf 6})$ & antiperiodic (1) & $({\bf 1},{\bf 1})_{X} +({\bf 1},{\bf 2})_{X+1/2} +({\bf 1},{\bf 3})_{X+1}$ & $1$ \\ \hline
	$Q_L^{}$ & & & $({\bf 3},{\bf 2})_{1/6}$ & $3$ \\ \hline
	$t_R^{}$ & & & $ ({\bf 3},{\bf 1})_{2/3}$ & $3$ \\ \hline
	$b_R^{}$ & & & $({\bf 3},{\bf 1})_{-1/3}$ & $3$ \\ \hline
\end{tabular}
\end{table}

For the sake of simplicity, we take $k_j=\tilde{k}_j^{}$, and
the counterparts in the gauge and Higgs sectors are set to the unity in the following analysis.
In place of $M_j$ and $e^a_i$, we introduce dimensionless parameters $\lambda^j =\pi R M_j^{}$ and $\epsilon_i^a=\sqrt{\pi R/2}e^a_i$ ($a=t,b$; $i=1,2$), respectively.
We are left with 10 input parameters, $k_t, k_b, k_A, \lambda^t, \lambda^b, \lambda^A, \epsilon^t_1, \epsilon^t_2, \epsilon^b_1$ and $\epsilon^b_2$.

\subsection{Mass Spectrum}
\label{sec:mass}

Here, we briefly describe the mass spectrum of this model. 
In the absence of 5D mass terms for the bulk gauge fields,
the masses of their Kaluza-Klein (KK) modes at the level $n$ scale as $m^{(n)}=n/R$.
When the EWSB takes place, the zero-mode and KK-mode masses 
of the gauge fields that are coupled with 
nonzero $\langle A_5^a \rangle$ are shifted upward by $\alpha$.
As for the $U(1)_X^{}$ gauge sector, due to quantum anomaly, 
the zero mode of $A_X^{}$ acquire a 
brane-localized mass $M_X^{}$, whose natural
value is the cut-off scale of the model~\cite{Panico:2005dh,Scrucca:2003ra}.
In the rest of this paper, we disregard $A_X^{}$ in the low-energy effective theory we discuss.
For details, see Refs.~\cite{Panico:2005dh,Scrucca:2003ra}.
In the limit where $M_X \to \infty$, 
the masses of the weak gauge bosons are given by
\begin{align}
m_{W^{\pm}}^{(n)}(\alpha)=\frac{n+\alpha}{R} ,\quad
m_Z^{(n)}(\alpha)=\frac{n+\alpha \sec\theta_W^{}}{R} .
\end{align}
Therefore, we can obtain $m_{W^{\pm}}^{(0)} = m_Z^{(0)} \cos \theta_W^{}$ for the zero modes.

The computation of the 4D mass spectrum for the bulk fermions is more complicated.
We consider the cases where 4D fermions arising from the bulk fermions are sufficiently heavy
and the effect of the bulk-brane mixing terms $\epsilon^a_i$ is negligible 
in computing the 4D masses of the bulk fermions. 
In this approximation, the 4D masses of the bulk fermions take the form of
\begin{align}
m_j^{(n)}(q\alpha)=\sqrt{\left(k_j\frac{n+\eta/2
+q\alpha}{R} \right)^2 +\left(\frac{\lambda^j}{\pi R} \right)^2},
\end{align}
where $q$ is the charge related to the coupling with the Higgs doublet, 
and determined from the $SU(3)_w^{}$ representation of the bulk fermion.
The fundamental representation with a $SU(3)_w^{}$ index can have $q=0,1$, and the symmetric representation with two $SU(3)_w^{}$ indices can have $q=0,1,2$.
The parameter $\eta=0 \,(1)$ stands for periodic (antiperiodic) bulk fermions.
Using this formalism, there appear five sets of KK masses that depends on $\alpha$:
$m_t^{(n)}(\alpha) , m_t^{(n)}(2\alpha) , m_b^{(n)}(\alpha) ,
m_A^{(n)}(\alpha)$ and $m_A^{(n)}(2\alpha)$ .

The masses of the brane fermions are induced by the bulk-brane mixing terms.
By diagonalizing the full kinetic operators for the bulk and brane fermions, 
we obtain the mass eigenvalues,
\begin{align}
	&m^t_0 =\frac{k_t}{\sqrt{2}\pi R}\frac{\epsilon^t_1 \epsilon^t_2}{k_t^2}\mathrm{Im}f_1 (\lambda^t/k_t ,2\alpha),
	\nonumber
	\\
	&m^b_0 =\frac{k_b}{\pi R}\frac{\epsilon^b_1 \epsilon^b_2}{k_b^2}\mathrm{Im}f_1 (\lambda^b/k_b ,\alpha).
\end{align}
After rescaling the mass eigenvalues using the normalization factors,
\begin{align}
	&Z^t_i =1
	+\delta_{i1}\frac{(\epsilon^{b}_1)^2}{k_b\lambda^b}\mathrm{Re}f_0 (\lambda^b/k_b ,0)
	+\delta_{i2}\frac{(\epsilon^{t}_2)^2}{2k_t\lambda^t}\mathrm{Re}f_0 (\lambda^t/k_t ,0)
	+\frac{(\epsilon^{t}_i)^2}{2^{\delta_{i2}}k_t\lambda^t}\mathrm{Re}f_0 (\lambda^t/k_t ,2\alpha),
	\nonumber
	\\
	&Z^b_i =1
	+\frac{(\epsilon^{b}_i)^2}{k_b\lambda^b}\mathrm{Re}f_0 (\lambda^b/k_b ,\alpha)
	+\delta_{i1}\frac{(\epsilon^{t}_1)^2}{k_t\lambda^t}\mathrm{Re}f_0 (\lambda^t/k_t ,\alpha),
\end{align}
with
\begin{align}
	f_{\delta}(x,\alpha)=\sum^{\infty}_{k=-\infty}\mathrm{e}^{-|2k+\delta |(x+i\pi \alpha)},
\end{align}
we obtain the physical masses of the brane fermions as
\begin{align}
m^a=\frac{m_0^a}{\sqrt{Z_1^a Z_2^a}} .
\label{eq:brane}
\end{align}

\section{Effective Potential}
\label{sec:Veff}

As a remnant of 5D gauge invariance, the Higgs potential is zero at the leading order
of perturbation theory.
However, the Higgs potential is induced by radiative corrections.
The effective potential at the one-loop level is given by
\begin{align}
V_{\rm eff}(\alpha)= \sum_I \frac{\sigma_I}{2} \int \frac{d^4p_E}{(2\pi)^4} \ln 
\left\{ p_E^2+m_I(\alpha)^2 \right\}  ,
\end{align}
where $\sigma_I^{}=1$ for bosons and $\sigma_I^{}=-1$ for fermions, and
the sum runs over all 4D fields whose masses depend on $\alpha$.
As for the 4D fields originating from the bulk fields, 
we perform the integration of the Euclidean momentum $p_E^{}$ for the KK modes
and decompose these contributions into an infinite $\alpha$-independent part and finite 
$\alpha$-dependent part~\cite{Antoniadis:2001cv}.
After the integration, the one-loop contribution from each bulk gauge field takes the form of
\begin{align}
V_V^{}(\alpha)=-\frac{9}{64\pi^6 R^4}\sum_{n=1}^{\infty}\frac{1}{n^5}\cos(2\pi n \alpha) .
\label{eq:Vg}
\end{align}
Then, the total contribution from the bulk gauge fields is given by
\begin{align}
V_g^{}(\alpha)=2V_V^{}(\alpha)+V_V^{}(\alpha \sec\theta_W^{} ) .
\end{align}

Similarly, we can calculate contributions from the bulk fermion pairs to the effective potential.
Taking the effects of the 5D mass parameter $\lambda^j$ and
the Lorentz-violating parameter $k_j^{}$, 
the contribution from $(\Psi_j^{},\bar{\Psi}_j^{})$ pair is given by
\begin{align}
V_{\Psi_j}(q\alpha)=
 \frac{3k_j^4 C_F^{}}{8\pi^6R^4}\sum_{n=1}^{\infty}\frac{(\sigma_S)^n}{n^5}
\left[1+2n\frac{\lambda^j}{k_j^{}}+\frac{4}{3}n^2\frac{(\lambda^j)^2}{k_j^2} \right]
\mathrm{e}^{-2n \lambda^j /k_j^{}}\cos(2\pi n q\alpha )  ,
\label{eq:Vf}
\end{align}
where $\sigma_S^{}= (-1)^\eta$ and $C_{F}^{}$ stands for the color factor.
Then, the total contribution from the bulk fermion pairs is given by
\begin{align}
V_f(\alpha)=V_{\Psi_t}(\alpha)+V_{\Psi_t}(2\alpha)+V_{\Psi_b}(\alpha)+V_{\Psi_A}(\alpha)+V_{\Psi_A}(2\alpha) .
\end{align}

As for the brane fermions, we need to incorporate the normalization factors
of the kinetic operators as well as the mass eigenvalues given in Eq.~(\ref{eq:brane}) into the calculation of the effective potential.
By changing the integral variable from $p_E^{}$ to $x=\pi R p_E^{}$,
one obtains
\begin{align}
V_t^{}(\alpha)=&\frac{-C_{F}^{}}{4\pi^6R^4}\int_0^{\infty} d x \,x^3\ln
\left[ \prod^2_{i=1} \mathrm{Re}\left[1
+\delta_{i1}\frac{(\epsilon^{b}_1)^2}{k_b x^b}f_0^{} \left(\frac{x^b}{k_b},0 \right)
+\delta_{i2}\frac{(\epsilon^{t}_2)^2}{2k_t x^t}f_0^{} \left(\frac{x^t}{k_t},0 \right)
\right. \right.
\nonumber
\\
&\left. \left.
+\frac{(\epsilon^{t}_1)^2}{2^{\delta_{i2}}k_t x^t}f_0 \left(\frac{x^t}{k_t},2\alpha \right) \right]
+\prod^2_{i=1} \mathrm{Im}
\frac{(\epsilon^{t}_i)^2}{2^{\delta_{i2}}k_t x} \left[f_1 \left(\frac{x^t}{k_t},2\alpha \right) \right]
\right] ,
\end{align}
and
\begin{align}
V_b(\alpha)=&\frac{-C_{F}^{}}{4\pi^6R^4}\int_0^{\infty}dx \,x^3\ln
\left[ \prod^2_{i=1} \mathrm{Re}\left[1
+\frac{(\epsilon^{b}_i)^2}{k_b x^b}f_0 \left(\frac{x^b}{k_b},\alpha \right)
+\delta_{i1}\frac{(\epsilon^{t}_1)^2}{k_t x^t}f_0 \left(\frac{x^t}{k_t},\alpha \right) \right]
\right.
\nonumber
\\
&\left.
+\prod^2_{i=1}
\mathrm{Im}\left[\frac{(\epsilon^{b}_i)^2}{k_b x} f_1 \left(\frac{x^b}{k_b},\alpha \right) \right]
\right] ,
\end{align}
with
\begin{align}
x^a=\pi R\sqrt{p_E^2+M_a^2}.
\end{align}
These contributions also turn out to be finite up to
the $\alpha$-independent constant part as in the case of the bulk fields.

Putting the above contributions together, 
we obtain the total one-loop effective potential,
\begin{align}
  V_{\rm eff}^{}(\alpha)=V_g^{}(\alpha) +V_f^{}(\alpha) +V_t^{}(\alpha) +V_b^{}(\alpha).
\end{align}
The EWSB occurs when 
$\alpha$ develops a non-zero VEV $\alpha_0^{}$.
The value of $\alpha_0^{}$ is obtained by the tadpole condition,
\begin{align}
\left(\frac{g R}{2}\right) \left.\frac{\partial V_{\rm eff}}{\partial \alpha}\right|_{\alpha=\alpha_0}=0.
\end{align}
The mass squared of the Higgs boson is given by
\begin{align}
m_h^2=\left(\frac{g R}{2} \right)^2
\left. \frac{\partial^2 V_{\rm eff}(\alpha)}{\partial \alpha^2}\right|_{\alpha=\alpha_0}  ,
\end{align}
and the triple Higgs boson coupling by
\begin{align}
\lambda_{hhh}^{}=\left(\frac{g R}{2} \right)^3
\left. \frac{\partial^3 V_{\rm eff}(\alpha)}{\partial \alpha^3}\right|_{\alpha=\alpha_0} .
\end{align}
The compactification scale $1/R$ is determined by using the relation $m_{W^{\pm}}^{(0)}=\alpha_0^{}/R$.
In order to avoid experimental constraints, $1/R$ should be sufficiently
large. In other words, $\alpha_0^{}$ should be very small.
It should be emphasized that tuning of model parameters is required 
for obtaining very small $\alpha_0^{}$.
This is because the Higgs potential in GHU is generated only
by radiative corrections and is sensitive to model parameters.
Therefore, it is nontrivial whether the Higgs field
develops a nonzero VEV.
We also have to adjust the mass of the Higgs boson to its experimental 
value by tuning the model parameters.
Namely, the mass of the discovered Higgs boson and the tadpole condition strongly constrain the shape of the Higgs potential and the allowed model parameter space.

\section{Constraints}
\label{sec:constraints}

Before discussing the triple Higgs boson coupling, 
we mention several constraints on the model parameters.
The experimental limits on 
the masses of the top quark and the Higgs boson, and the compactification scale put severe constraints on GHU models in generic.
In the previous study performed in Ref.~\cite{Panico:2005dh}, 
the Lorentz-violating parameters $k_j$ are introduced 
to adjust the masses of the top quark and the Higgs boson.
In Refs.~\cite{Panico:2005dh,Panico:2006em}, the $\rho$ parameter and the $Zb_L^{}\bar{b}_L^{}$ coupling are investigated and turned out to be strongly affected by the KK modes of the bulk fields, such as the sine modes of $A_X^{(n)}$ and the lightest modes of the bulk fermions.
An improved analysis of electroweak observables in a slightly refined $SU(3)_w^{}$ GHU model puts the lower bound on the compactification scale as $1/R \gtrsim 5$~TeV for $m_h^{}=125$~GeV~\cite{Panico:2006em}.
Since the works of Refs.~\cite{Panico:2005dh,Panico:2006em} were done
before the discovery of the Higgs boson, 
we revisit the $SU(3)_w^{}$ model to investigate viable parameter regions.

\begin{figure}[t]
  \begin{center}
    \begin{tabular}{c}

      \begin{minipage}{0.45\hsize}
        \begin{center}
          \includegraphics[clip, width=7cm]{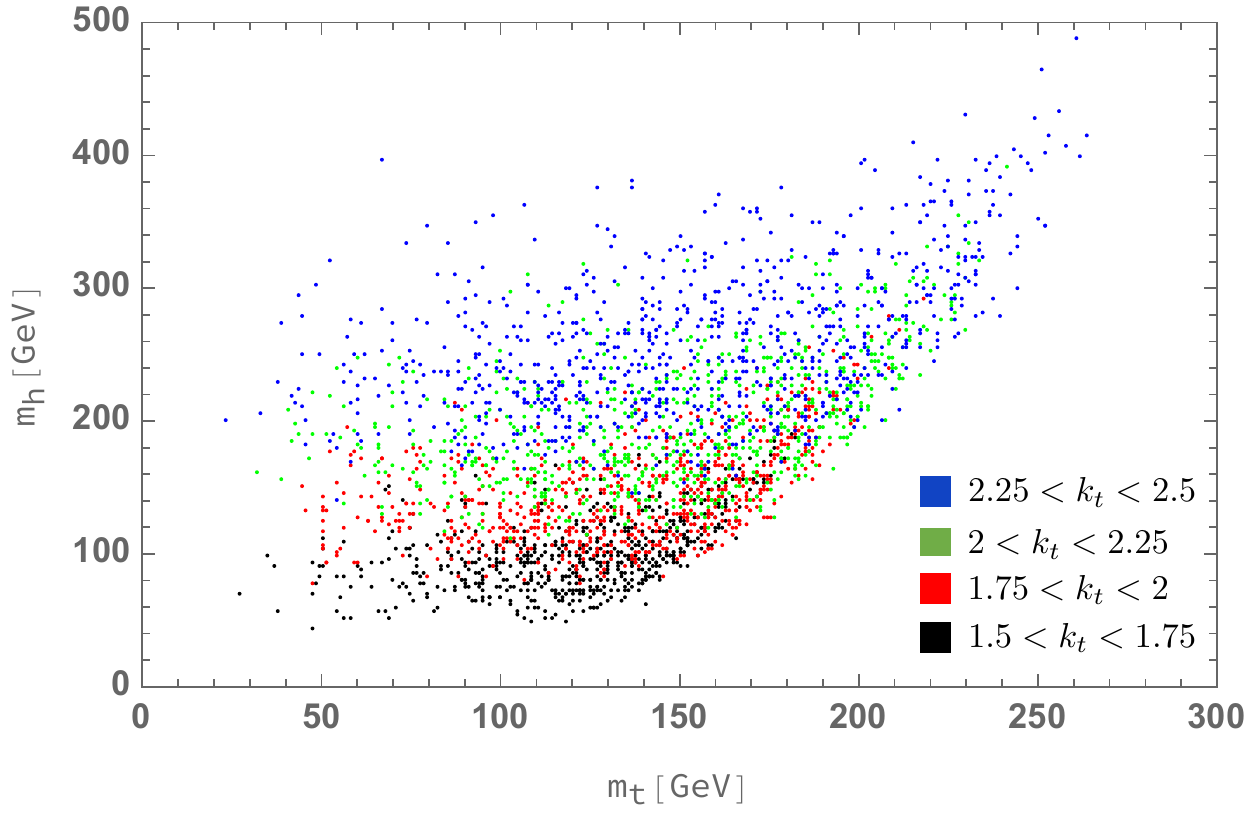}
        \end{center}
      \end{minipage}

      \begin{minipage}{0.45\hsize}
        \begin{center}
          \includegraphics[clip, width=7cm]{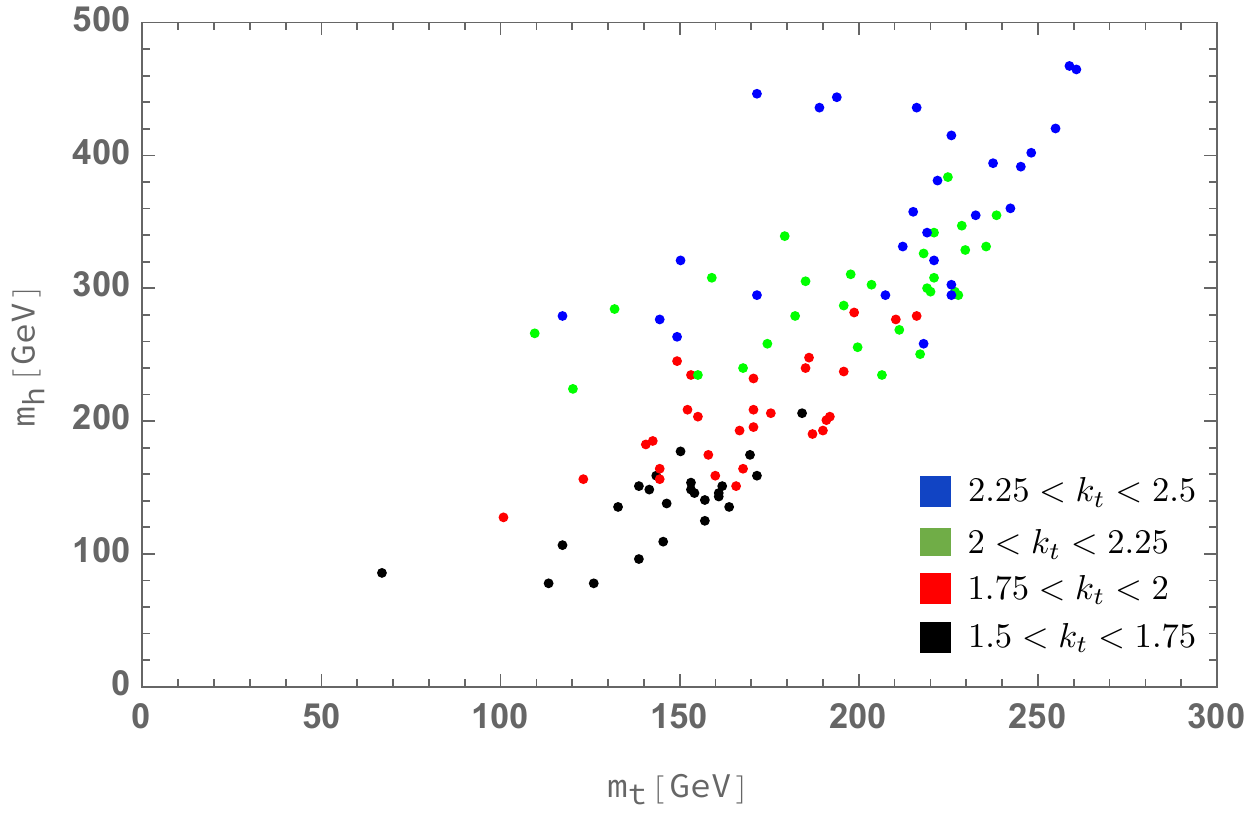}
        \end{center}
      \end{minipage}

    \end{tabular}
    \caption{Scatter plots of the predicted values of the top quark mass $m_t^{}$ and the Higgs
    boson mass $m_h^{}$ for $1/R>800$ GeV (left) and for $1/R>5$ TeV (right).
    Color codes are assigned to each range of the Lorentz violating parameter $k_t^{}$:
Blue points stands for $2.25<k_t^{}<2.5$, green for $2<k_t^{}<2.25$, red for $1.75<k_t^{}<2$,
and black for $1.5<k_t^{}<1.75$.}
    \label{fig:const}
  \end{center}
\end{figure}

We perform a random scan of model parameters,
and numerically evaluate the masses of the top quark and the Higgs boson as well as the compactification scale from the effective potential.
We scan the following model parameter regions in accordance with the previous study~\cite{Panico:2005dh}:
\begin{align}
\left\{
\begin{array}{l}
	1.5 <k_t^{} < 2.5, \\
	1.25 <k_b^{} < 2.25, \\
	1.1\times k_t^{} < k_A < 1.5 \times k_t^{},
\end{array}
\right. \hspace{5mm} \left\{
\begin{array}{l}
	0.5 <\lambda^t < 1.5, \\
	5 <\lambda^b < 7, \\
	0.75 < \lambda^A < 3.5,
\end{array}
\right. \hspace{5mm} \left\{
\begin{array}{l}
	0.75 <\epsilon^t_{1,2} < 7.5, \\
	2 <\epsilon^b_{1,2} < 7.
\end{array}
\right.
\label{eq:parameter}
\end{align}
Figure~\ref{fig:const} shows the scatter plots of
the predicted values of the top quark mass $m_t^{}$ and the Higgs boson mass $m_h^{}$.
The left (right) panel shows cases
for compactification scales larger than $800~\mathrm{GeV}$ ($5~\mathrm{TeV}$).
Color codes are assigned to each range of the Lorentz violating parameter $k_t^{}$:
Blue points stands for $2.25<k_t^{}<2.5$, green for $2<k_t^{}<2.25$, red for $1.75<k_t^{}<2$,
and black for $1.5<k_t^{}<1.75$.
The left plot is consistent with the results presented in Ref.~\cite{Panico:2005dh}.
It is clear from the plots that 
the mass of the Higgs boson is positively correlated with $k_t^{}$.
As the compactification scale $1/R$ is increased, a finer tuning is required for obtaining the correct weak-scale VEV.
Therefore, as shown in the right panel of Fig.~\ref{fig:const}, 
the number of allowed parameter sets is smaller 
if we impose the experimental lower bound on the compactification scale,
$1/R>5~\mathrm{TeV}$~\cite{Panico:2006em}.
The bottom line is that there is still enough room for 
reproducing the measured values of the masses of the top and Higgs boson
for $1.5 <k_t^{} <2$.

\section{Triple Higgs Boson Coupling}
\label{sec:hhh}

After imposing the above-mentioned experimental constraints, 
we investigate to what extent the predicted triple Higgs boson coupling $\lambda_{hhh}^{}$
can deviate from the corresponding SM value $\lambda_{hhh}^{\rm SM}$.
The current experimental constraint on $\lambda_{hhh}^{}$ is still weak.
Searches for Higgs boson pair production at the LHC have put constraints on
the triple Higgs boson coupling as $-5.0 <\lambda_{hhh}^{}/\lambda_{hhh}^{\rm SM} <12.0$ for ATLAS~\cite{Aaboud:2017vzb} and 
$-11.8 <\lambda_{hhh}^{}/\lambda_{hhh}^{\rm SM} <18.8$ for 
CMS~\cite{Sirunyan:2018koj} at the 95\% confidence level.
At future collider experiments,
these bounds will be significantly improved.
At the High-Luminosity LHC,
$0.52 < \lambda_{hhh}^{}/\lambda_{hhh}^{\rm SM} < 1.5$ and $0.57< \lambda_{hhh}^{}/\lambda_{hhh}^{\rm SM} < 1.5$ with and without systematic uncertainties ($1~\sigma$), respectively~\cite{Cepeda:2019klc}.
At the ILC with a center-of-mass energy of $\sqrt{s}=1~\mathrm{TeV}$ and integrated luminosity of $L=4~{\rm ab}^{-1}$, a precision of 10\% is estimated~\cite{Fujii:2019zll}.
At the CLIC with $\sqrt{s}=3$~TeV and $L=5~{\rm ab}^{-1}$, the triple Higgs boson coupling will be measured with a relative uncertainty of $-8\%$ to $11\%$ ($1~\sigma$)~\cite{Roloff:2019crr}.

\subsection{Numerical Result}

\begin{figure}[t]
 	\begin{center}
		\includegraphics[clip,width=9cm]{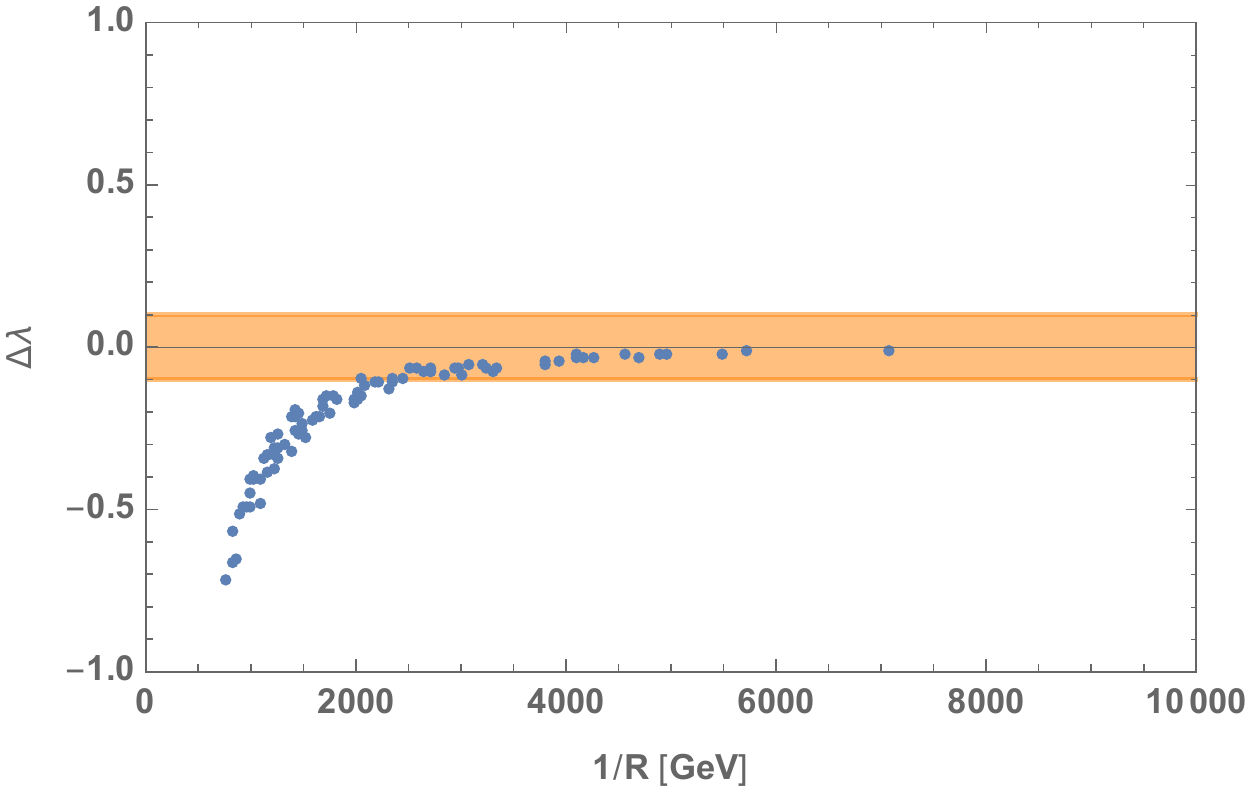}
		\caption{Correlation between the deviation of the triple Higgs boson coupling $\Delta \lambda$
		and the compactification scale $1/R$. The orange band shows the $1~\sigma$ accuracy expected at the ILC.}
		\label{fig:Dl}
	\end{center}
\end{figure}

As a measure of the triple Higgs boson coupling,
we introduce the deviation parameter defined by
\begin{align}
\Delta\lambda=\frac{\lambda_{hhh}^{}-\lambda_{hhh}^{\rm SM}}{\lambda_{hhh}^{\rm SM}} .
\end{align}
In evaluating ${\lambda_{hhh}^{\rm SM}}$,
we use the effective potential method and include the top quark contribution 
at the one-loop level.
Actually, the evaluated values of the masses of the top quark and Higgs boson
and other 4D parameters should be regarded as 
those at some high energy scale, and 
the corresponding measured values are corrected using renormalization
group equations.
Detailed analysis of the renormalization group equations is
beyond the scope of this paper.
Taking such uncertainties into account,
we employ $152~{\rm GeV} <m_t <182~{\rm GeV}$ and $110~{\rm GeV} <m_h <140~{\rm GeV}$
as allowed values in our parameter scan, whose range is given in Eq.~(\ref{eq:parameter}).
In Fig.~\ref{fig:Dl},
we show the correlation between the deviation of the triple Higgs boson coupling $\Delta \lambda$ and 
the compactification scale $1/R$.
The orange band shows the $1~\sigma$ accuracy expected at the ILC.
From Fig.~\ref{fig:Dl}, it is clear that the deviation $\Delta\lambda$ 
is primarily characterized by the compactification scale $1/R$.
At very large compactification scales, the deviation of the triple Higgs boson coupling almost vanishes as
\begin{align}
\lim_{1/R \to \infty} \Delta\lambda =0.
\label{eq:lambda}
\end{align}
Compared with the current constraint on the compactification scale, $1/R>5~\mathrm{TeV}$,  additional extensions to this GHU model are unavoidable
if a significant deviation is observed at the future colliders.

\begin{figure}[t]
  \begin{center}
    \begin{tabular}{c}

      \begin{minipage}{0.45\hsize}
        \begin{center}
          \includegraphics[clip, width=7cm]{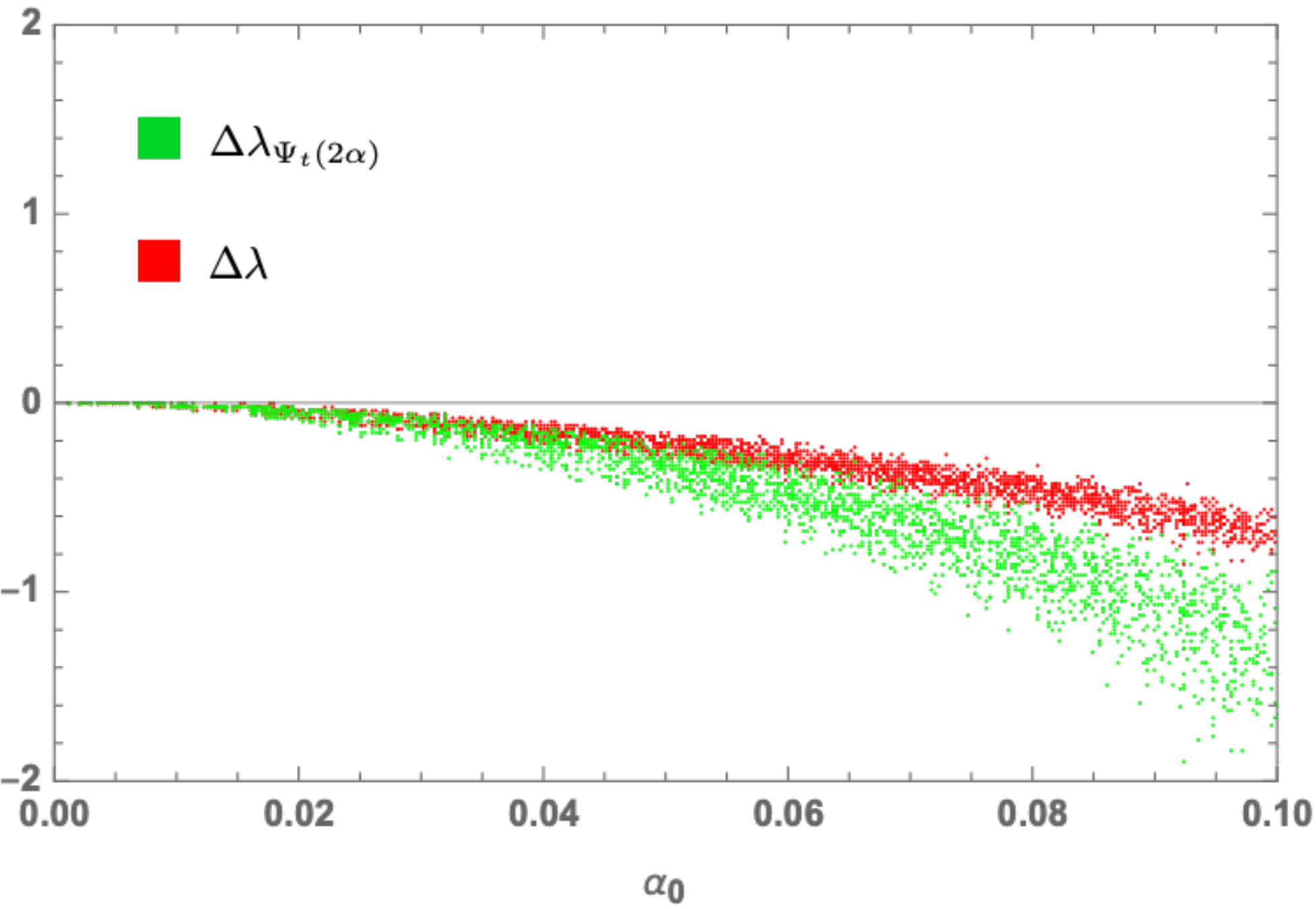}
        \end{center}
      \end{minipage}

      \begin{minipage}{0.45\hsize}
        \begin{center}
          \includegraphics[clip, width=7cm]{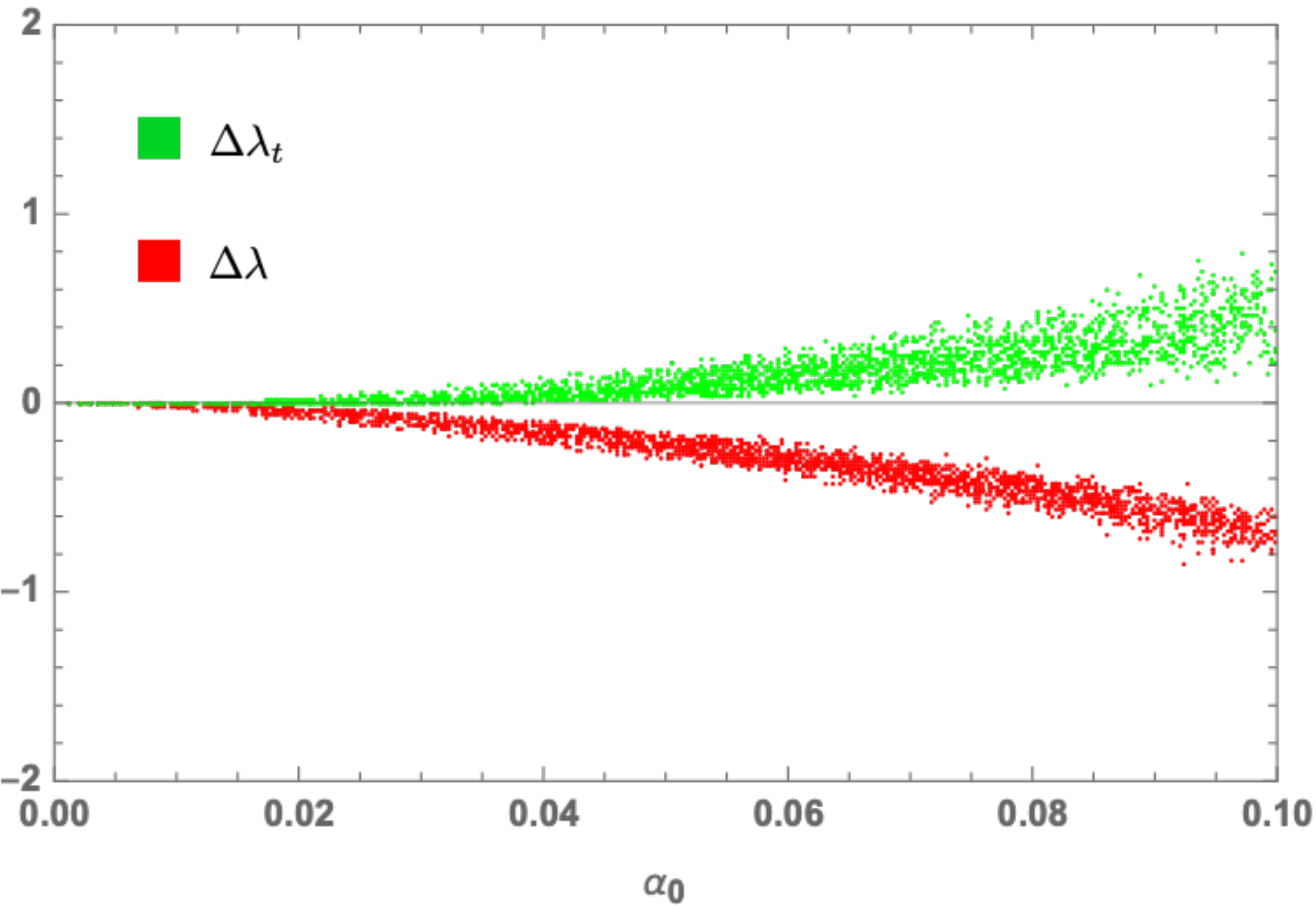}
        \end{center}
      \end{minipage}
      
      \\
      
      \begin{minipage}{0.45\hsize}
        \begin{center}
          \includegraphics[clip, width=7cm]{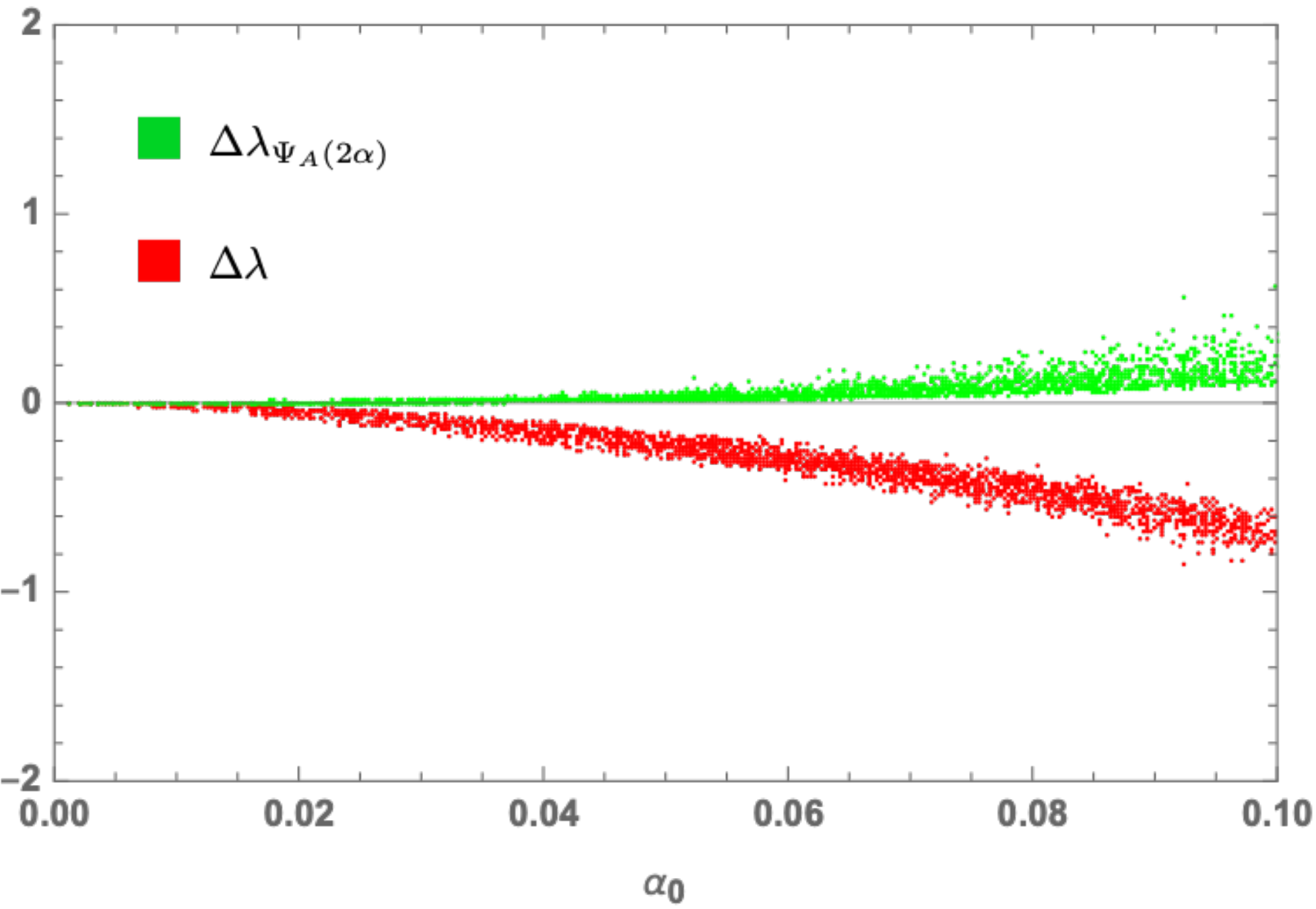}
        \end{center}
      \end{minipage}

      \begin{minipage}{0.45\hsize}
        \begin{center}
          \includegraphics[clip, width=7cm]{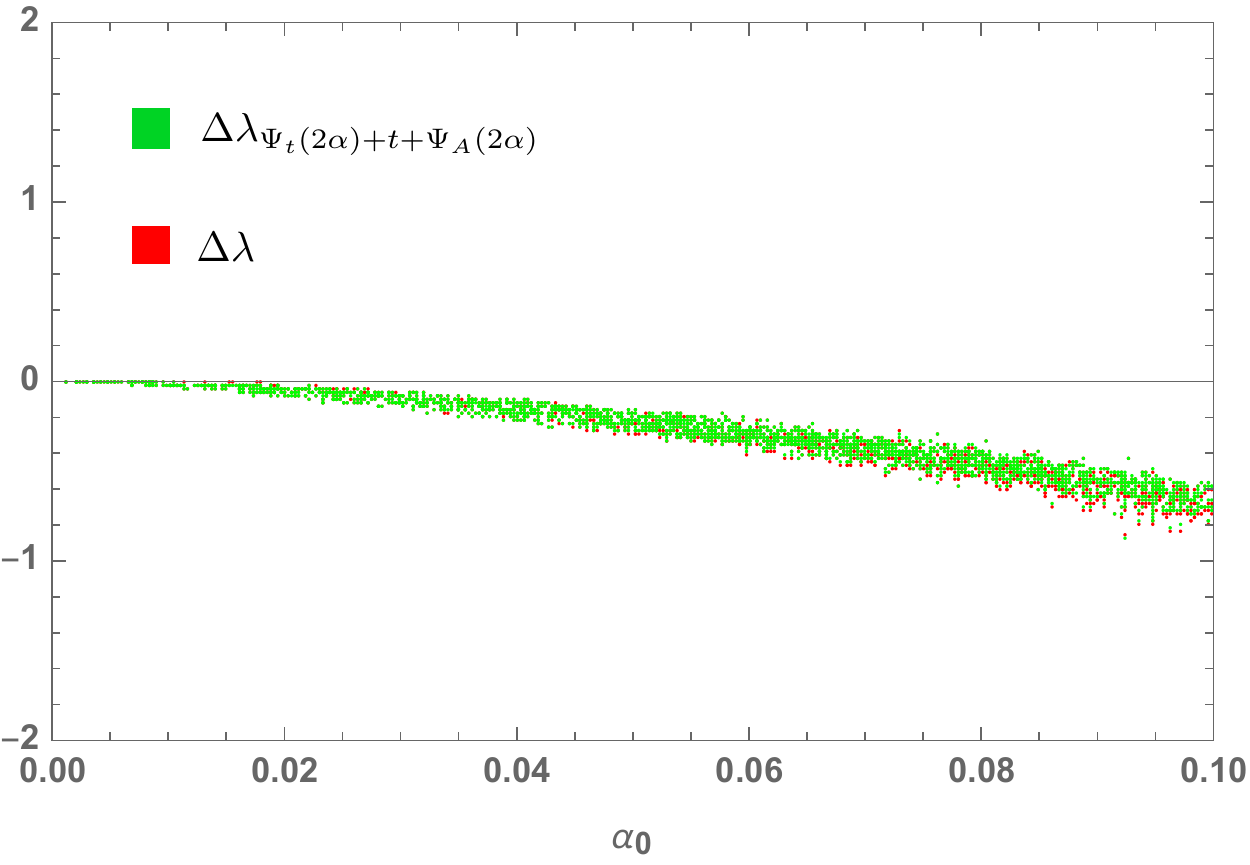}
        \end{center}
      \end{minipage}

    \end{tabular}
    \caption{Contributions to the deviation parameters 
$\Delta \lambda_i^{}$ for $\Psi_{t}(2\alpha)$ (upper left panel), $t$ (upper right), $\Psi_{A}(2\alpha)$ (lower left), and their sum (lower right).
For comparison, the total deviation parameter $\Delta \lambda$ is also displayed
(red).}
    \label{fig:con}
  \end{center}
\end{figure}

Let us investigate to what extent each field contributes to the total deviation of the triple Higgs boson coupling $\Delta \lambda$.
To this end, we decompose the predicted triple Higgs boson coupling into potential
terms as
\begin{align}
    \lambda_{hhh}^{} = \sum_i \delta \lambda_i^{},
\end{align}
where $\delta \lambda_i^{}$ denotes the contribution from a field $i$,
\begin{align}
\quad \delta \lambda_i = \left. \frac{\partial^3 V_i}{\partial \alpha^3}\right|_{\alpha=\alpha_0} -
\left.\frac{3}{\alpha}\frac{\partial^2 V_i}{\partial \alpha^2}\right|_{\alpha=\alpha_0}
+\left.\frac{3}{\alpha^2}\frac{\partial V_i}{\partial \alpha}\right|_{\alpha=\alpha_0}.
\end{align}
It is convenient to introduce the following deviation parameter:
\begin{align}
	\Delta \lambda_i =\frac{\delta \lambda_i-\lambda_{hhh}^{\rm SM}}{\lambda_{hhh}^{\rm SM}},
\end{align}
which quantifies the extent of the contribution from the field $i$.
In Fig.~\ref{fig:con}, green points shows the contributions to the deviation parameters 
$\Delta \lambda_i^{}$ for $\Psi_{t}^{}(2\alpha)$ (upper left panel), $t$ (upper right), $\Psi_{A}^{}(2\alpha)$ (lower left), and their sum (lower right),
\begin{align}
	\Delta \lambda_{\Psi_{t}^{}(2\alpha)+t+\Psi_{A}^{}(2\alpha)} 
	=\frac{\delta \lambda_{\Psi_{t}(2\alpha)}^{}
	+ \delta \lambda_t + \delta \lambda_{\Psi_{A}(2\alpha)}^{}-\lambda_{hhh}^{\rm SM}}{\lambda_{hhh}^{\rm SM}}.
\end{align}
For comparison, the total deviation parameter $\Delta \lambda$ is also displayed
(red).
Particles with larger $q$ values have stronger couplings to the Higgs field, and give dominant contributions
to the deviation of the triple Higgs boson coupling.
The total deviation $\Delta \lambda$ is well approximated by 
the sum of the contributions from $\Psi_{t}(2\alpha)$, $t$ and $\Psi_{A}(2\alpha)$.

\subsection{Analysis}

From the numerical results presented above, the deviation of the triple Higgs boson coupling converges to zero as the compactification scale increases.
This means that the shape of our GHU Higgs potential resembles the SM Higgs potential
at around the VEV for small $\alpha_0^{}$ although their tree-level shapes are completely different.
Here, we analyze the shape of the Higgs potential at small $\alpha$.

In general, the Higgs potential in GHU with a flat extra dimension is constructed by summing up the contributions of bulk gauge fields, bulk fermions, and brane fermions, as discussed in Sec.~\ref{sec:Veff}.
Although the field-dependent masses of the brane fermions are complicated functions of $\alpha$ as discussed in Sec.~\ref{sec:mass},
they are approximated by linear functions of $\alpha$ for $\alpha \ll 1$ as is the case for the fermions in the SM.
Expanding the GHU Higgs potential with respect to $\alpha$, we obtain
\begin{align}
V_{\rm eff}(\phi) \simeq - \left(\frac{1}{R} \right)^2 A \phi^2 +B\phi^4 +C\phi^4\ln\frac{\phi^2}{v^2} +\left(\frac{1}{R} \right)^{-2}D\phi^6 ,
\label{eq:Vapp}
\end{align}
where $A$, $B$, $C$ and $D$ are dimensionless parameters that are functions of the model parameters.
From the tadpole condition,
\begin{align}
\left.\frac{\partial V_{\rm eff}}{\partial \phi}\right|_{\phi=v} =0,
\end{align}
and the condition for reproducing the mass of the discovered Higgs boson,
\begin{align}
\left. \frac{\partial^2 V_{\rm eff}}{\partial \phi^2}\right|_{\phi=v} =m_h^2,
\end{align}
we obtain
\begin{align}
 0 = -\left(\frac{1}{R} \right)^2 \frac{A}{2v^2} +B + \frac{C}{2} + \frac{3}{2}\left(\frac{1}{R} \right)^{-2}D v^2,
 \label{eq:tadpole}
\end{align}
and
\begin{align}
m_h^2 =4\left(\frac{1}{R} \right)^2 A +8C v^2 +12\left(\frac{1}{R} \right)^{-2} Dv^4,
\label{eq:higgsmass}
\end{align}
respectively. Namely, two of the four coefficients are not independent.
Since the compactification scale should be above a few TeV ($\alpha_0^{} \ll 1$),
Eqs.~(\ref{eq:tadpole}) and (\ref{eq:higgsmass}) suggests that the parameter $A$ should be
very small for accounting for the VEV and the mass of the Higgs boson.
This fact implies the so-called little hierarchy problem.
In GHU, one needs to tune bulk field parameters to adequately suppress the parameter $A$.
In our GHU model, this suppression is accomplished by introducing the additional bulk fields
($\Psi_A^{}, \tilde{\Psi}_A^{}$).

Eliminating the parameters $A$ and $B$ using the tadpole and mass conditions, 
Eqs.~(\ref{eq:tadpole}) and (\ref{eq:higgsmass}),
the triple Higgs boson coupling is written as 
\begin{align}
\lambda_{hhh}=\frac{3}{v}\left[ m_h^2 +\frac{16}{3}Cv^2 +16\left(\frac{1}{R} \right)^{-2} Dv^4 \right] .
\end{align} 
Since the term proportional to $\phi^4 \ln \phi^2$ arises from 
loop diagrams in which only SM particles are involved,
the value of the coefficient $C$ is the same as that in the SM.
Therefore, the deviation of the triple Higgs boson coupling from the SM is 
suppressed by the square of the compactification scale as
\begin{align}
	\Delta \lambda = \frac{48 Dv^3}{\lambda_{hhh}^{\rm SM}} \left( \frac{1}{R} \right)^{-2}.
	\label{eq:Deltalambda}
\end{align}
Namely, in the cases when the $\phi^6$ term and higher order ones are negligibly small,
the shape of the potential and the triple Higgs boson coupling
should be the same as the SM ones irrespective of the origin of the potential.
In deriving Eq.~(\ref{eq:Deltalambda}), we have not assumed specific bulk fields or Lorentz-violating parameters.
We emphasize that such a conclusion is applicable to a wide range of GHU models 
that reduces to the Higgs sector with one Higgs doublet below the compactification scale of a flat extra dimension.
This point has not been manifested in earlier works.
Therefore, if a significant deviation of the triple Higgs boson coupling 
is observed at collider experiments in the future,
we have to extensively modify the GHU framework.

Finally, we mention the sign of the deviation parameter, which is controlled by the coefficient $D$.
As for the bulk fields, the deviation derived from each contribution tends to take a larger value when $p$ and $k$ are large and $\lambda$ is small. 
Its sign is given by $\sigma_S \sigma_I$, which is 
determined by the periodicity on $S^1$ and the spin of each bulk field. 
In our model, the deviation goes in the negative direction because the bulk fermion $\Psi(2\alpha)$ with $\sigma_S^{}=1$ and $\sigma_I^{}=-1$ gives rise to the largest contribution. 
This observation about the dependence on $R$ and the sign  accounts for our numerical results about the
deviation of the triple Higgs boson coupling presented in Fig.~\ref{fig:Dl}.

\section{Summary}
\label{sec:summary}

In sharp contrast to the Higgs potential assumed in the SM,
the GHU Higgs potential is formed exclusively from quantum corrections 
because the leading-order Higgs potential vanishes as a remnant of higher-dimensional gauge symmetry.
In order to investigate phenomenological difference between the GHU Higgs potential and SM one,
we have focused on the triple Higgs boson coupling.
Taking the $SU(3)_w^{}$ model with 5D Lorentz symmetry relaxed
as an example of the GHU framework with a flat extra dimension,
it has been illustrated that the deviation of the triple Higgs boson coupling from the SM prediction
quickly vanishes as the compactification scale $1/R$ increases.
For compactification scales larger than the experimental limit $1/R \gtrsim 5~\mathrm{TeV}$,
the deviation has been found to be smaller than 10\%.

We have also discussed the characteristics of the Higgs potential in GHU 
in which one Higgs doublet appears below the compactification scale of a flat extra dimension.
We have expanded the Higgs potential with respect to $\alpha$,
and found that the GHU potential has the same shape 
as the SM one up to higher-order terms suppressed by the compactification scale.
This fact indicates that the deviation of the triple Higgs boson coupling vanishes at large compactification scales in such a class of GHU models.
Therefore, compared with the expected accuracy of the measurement of the triple Higgs boson coupling, 
the observation of its noticeable deviation at future collider experiments
will demand us to significantly modify the GHU framework.

\begin{acknowledgments}
The authors would like to thank Shuichiro Funatsu for useful discussion and comments.
The work of M. K. was supported, in part, by JSPS Grant-in-Aid for Scientific Research
KAKENHI Grant Numbers 20H00160 and 21K03571.
The work of S. S. was supported, in part, by the Sasakawa Scientific Research Grant from The Japan Science Society.
\end{acknowledgments}

\bibliographystyle{apsrev4-1}

\bibliography{reference}

\begin{thebibliography}{29}%
\makeatletter
\providecommand \@ifxundefined [1]{%
 \@ifx{#1\undefined}
}%
\providecommand \@ifnum [1]{%
 \ifnum #1\expandafter \@firstoftwo
 \else \expandafter \@secondoftwo
 \fi
}%
\providecommand \@ifx [1]{%
 \ifx #1\expandafter \@firstoftwo
 \else \expandafter \@secondoftwo
 \fi
}%
\providecommand \natexlab [1]{#1}%
\providecommand \enquote  [1]{``#1''}%
\providecommand \bibnamefont  [1]{#1}%
\providecommand \bibfnamefont [1]{#1}%
\providecommand \citenamefont [1]{#1}%
\providecommand \href@noop [0]{\@secondoftwo}%
\providecommand \href [0]{\begingroup \@sanitize@url \@href}%
\providecommand \@href[1]{\@@startlink{#1}\@@href}%
\providecommand \@@href[1]{\endgroup#1\@@endlink}%
\providecommand \@sanitize@url [0]{\catcode `\\12\catcode `\$12\catcode
  `\&12\catcode `\#12\catcode `\^12\catcode `\_12\catcode `\%12\relax}%
\providecommand \@@startlink[1]{}%
\providecommand \@@endlink[0]{}%
\providecommand \url  [0]{\begingroup\@sanitize@url \@url }%
\providecommand \@url [1]{\endgroup\@href {#1}{\urlprefix }}%
\providecommand \urlprefix  [0]{URL }%
\providecommand \Eprint [0]{\href }%
\providecommand \doibase [0]{http://dx.doi.org/}%
\providecommand \selectlanguage [0]{\@gobble}%
\providecommand \bibinfo  [0]{\@secondoftwo}%
\providecommand \bibfield  [0]{\@secondoftwo}%
\providecommand \translation [1]{[#1]}%
\providecommand \BibitemOpen [0]{}%
\providecommand \bibitemStop [0]{}%
\providecommand \bibitemNoStop [0]{.\EOS\space}%
\providecommand \EOS [0]{\spacefactor3000\relax}%
\providecommand \BibitemShut  [1]{\csname bibitem#1\endcsname}%
\let\auto@bib@innerbib\@empty
\bibitem [{\citenamefont {Aad}\ \emph {et~al.}(2012)\citenamefont {Aad} \emph
  {et~al.}}]{Aad:2012tfa}%
  \BibitemOpen
  \bibfield  {author} {\bibinfo {author} {\bibfnamefont {G.}~\bibnamefont
  {Aad}} \emph {et~al.} (\bibinfo {collaboration} {ATLAS}),\ }\href {\doibase
  10.1016/j.physletb.2012.08.020} {\bibfield  {journal} {\bibinfo  {journal}
  {Phys. Lett. B}\ }\textbf {\bibinfo {volume} {716}},\ \bibinfo {pages} {1}
  (\bibinfo {year} {2012})},\ \Eprint {http://arxiv.org/abs/1207.7214}
  {arXiv:1207.7214 [hep-ex]} \BibitemShut {NoStop}%
\bibitem [{\citenamefont {Chatrchyan}\ \emph {et~al.}(2012)\citenamefont
  {Chatrchyan} \emph {et~al.}}]{Chatrchyan:2012ufa}%
  \BibitemOpen
  \bibfield  {author} {\bibinfo {author} {\bibfnamefont {S.}~\bibnamefont
  {Chatrchyan}} \emph {et~al.} (\bibinfo {collaboration} {CMS}),\ }\href
  {\doibase 10.1016/j.physletb.2012.08.021} {\bibfield  {journal} {\bibinfo
  {journal} {Phys. Lett. B}\ }\textbf {\bibinfo {volume} {716}},\ \bibinfo
  {pages} {30} (\bibinfo {year} {2012})},\ \Eprint
  {http://arxiv.org/abs/1207.7235} {arXiv:1207.7235 [hep-ex]} \BibitemShut
  {NoStop}%
\bibitem [{\citenamefont {Aad}\ \emph {et~al.}(2016)\citenamefont {Aad} \emph
  {et~al.}}]{Khachatryan:2016vau}%
  \BibitemOpen
  \bibfield  {author} {\bibinfo {author} {\bibfnamefont {G.}~\bibnamefont
  {Aad}} \emph {et~al.} (\bibinfo {collaboration} {ATLAS, CMS}),\ }\href
  {\doibase 10.1007/JHEP08(2016)045} {\bibfield  {journal} {\bibinfo  {journal}
  {JHEP}\ }\textbf {\bibinfo {volume} {08}},\ \bibinfo {pages} {045} (\bibinfo
  {year} {2016})},\ \Eprint {http://arxiv.org/abs/1606.02266} {arXiv:1606.02266
  [hep-ex]} \BibitemShut {NoStop}%
\bibitem [{\citenamefont {Aaboud}\ \emph {et~al.}(2018)\citenamefont {Aaboud}
  \emph {et~al.}}]{Aaboud:2017vzb}%
  \BibitemOpen
  \bibfield  {author} {\bibinfo {author} {\bibfnamefont {M.}~\bibnamefont
  {Aaboud}} \emph {et~al.} (\bibinfo {collaboration} {ATLAS}),\ }\href
  {\doibase 10.1007/JHEP03(2018)095} {\bibfield  {journal} {\bibinfo  {journal}
  {JHEP}\ }\textbf {\bibinfo {volume} {03}},\ \bibinfo {pages} {095} (\bibinfo
  {year} {2018})},\ \Eprint {http://arxiv.org/abs/1712.02304} {arXiv:1712.02304
  [hep-ex]} \BibitemShut {NoStop}%
\bibitem [{\citenamefont {Sirunyan}\ \emph {et~al.}(2019)\citenamefont
  {Sirunyan} \emph {et~al.}}]{Sirunyan:2018koj}%
  \BibitemOpen
  \bibfield  {author} {\bibinfo {author} {\bibfnamefont {A.~M.}\ \bibnamefont
  {Sirunyan}} \emph {et~al.} (\bibinfo {collaboration} {CMS}),\ }\href
  {\doibase 10.1140/epjc/s10052-019-6909-y} {\bibfield  {journal} {\bibinfo
  {journal} {Eur. Phys. J. C}\ }\textbf {\bibinfo {volume} {79}},\ \bibinfo
  {pages} {421} (\bibinfo {year} {2019})},\ \Eprint
  {http://arxiv.org/abs/1809.10733} {arXiv:1809.10733 [hep-ex]} \BibitemShut
  {NoStop}%
\bibitem [{\citenamefont {Wess}\ and\ \citenamefont
  {Zumino}(1974)}]{Wess:1974tw}%
  \BibitemOpen
  \bibfield  {author} {\bibinfo {author} {\bibfnamefont {J.}~\bibnamefont
  {Wess}}\ and\ \bibinfo {author} {\bibfnamefont {B.}~\bibnamefont {Zumino}},\
  }\href {\doibase 10.1016/0550-3213(74)90355-1} {\bibfield  {journal}
  {\bibinfo  {journal} {Nucl. Phys. B}\ }\textbf {\bibinfo {volume} {70}},\
  \bibinfo {pages} {39} (\bibinfo {year} {1974})}\BibitemShut {NoStop}%
\bibitem [{\citenamefont {Salam}\ and\ \citenamefont
  {Strathdee}(1974)}]{Salam:1974yz}%
  \BibitemOpen
  \bibfield  {author} {\bibinfo {author} {\bibfnamefont {A.}~\bibnamefont
  {Salam}}\ and\ \bibinfo {author} {\bibfnamefont {J.~A.}\ \bibnamefont
  {Strathdee}},\ }\href {\doibase 10.1016/0550-3213(74)90537-9} {\bibfield
  {journal} {\bibinfo  {journal} {Nucl. Phys. B}\ }\textbf {\bibinfo {volume}
  {76}},\ \bibinfo {pages} {477} (\bibinfo {year} {1974})}\BibitemShut
  {NoStop}%
\bibitem [{\citenamefont {Kaplan}\ and\ \citenamefont
  {Georgi}(1984)}]{Kaplan:1983fs}%
  \BibitemOpen
  \bibfield  {author} {\bibinfo {author} {\bibfnamefont {D.~B.}\ \bibnamefont
  {Kaplan}}\ and\ \bibinfo {author} {\bibfnamefont {H.}~\bibnamefont
  {Georgi}},\ }\href {\doibase 10.1016/0370-2693(84)91177-8} {\bibfield
  {journal} {\bibinfo  {journal} {Phys. Lett. B}\ }\textbf {\bibinfo {volume}
  {136}},\ \bibinfo {pages} {183} (\bibinfo {year} {1984})}\BibitemShut
  {NoStop}%
\bibitem [{\citenamefont {Manton}(1979)}]{Manton:1979kb}%
  \BibitemOpen
  \bibfield  {author} {\bibinfo {author} {\bibfnamefont {N.~S.}\ \bibnamefont
  {Manton}},\ }\href {\doibase 10.1016/0550-3213(79)90192-5} {\bibfield
  {journal} {\bibinfo  {journal} {Nucl. Phys. B}\ }\textbf {\bibinfo {volume}
  {158}},\ \bibinfo {pages} {141} (\bibinfo {year} {1979})}\BibitemShut
  {NoStop}%
\bibitem [{\citenamefont {Fairlie}(1979{\natexlab{a}})}]{Fairlie:1979zy}%
  \BibitemOpen
  \bibfield  {author} {\bibinfo {author} {\bibfnamefont {D.~B.}\ \bibnamefont
  {Fairlie}},\ }\href {\doibase 10.1088/0305-4616/5/4/002} {\bibfield
  {journal} {\bibinfo  {journal} {J. Phys. G}\ }\textbf {\bibinfo {volume}
  {5}},\ \bibinfo {pages} {L55} (\bibinfo {year}
  {1979}{\natexlab{a}})}\BibitemShut {NoStop}%
\bibitem [{\citenamefont {Fairlie}(1979{\natexlab{b}})}]{Fairlie:1979at}%
  \BibitemOpen
  \bibfield  {author} {\bibinfo {author} {\bibfnamefont {D.~B.}\ \bibnamefont
  {Fairlie}},\ }\href {\doibase 10.1016/0370-2693(79)90434-9} {\bibfield
  {journal} {\bibinfo  {journal} {Phys. Lett. B}\ }\textbf {\bibinfo {volume}
  {82}},\ \bibinfo {pages} {97} (\bibinfo {year}
  {1979}{\natexlab{b}})}\BibitemShut {NoStop}%
\bibitem [{\citenamefont {Hosotani}(1983{\natexlab{a}})}]{Hosotani:1983xw}%
  \BibitemOpen
  \bibfield  {author} {\bibinfo {author} {\bibfnamefont {Y.}~\bibnamefont
  {Hosotani}},\ }\href {\doibase 10.1016/0370-2693(83)90170-3} {\bibfield
  {journal} {\bibinfo  {journal} {Phys. Lett. B}\ }\textbf {\bibinfo {volume}
  {126}},\ \bibinfo {pages} {309} (\bibinfo {year}
  {1983}{\natexlab{a}})}\BibitemShut {NoStop}%
\bibitem [{\citenamefont {Hosotani}(1983{\natexlab{b}})}]{Hosotani:1983vn}%
  \BibitemOpen
  \bibfield  {author} {\bibinfo {author} {\bibfnamefont {Y.}~\bibnamefont
  {Hosotani}},\ }\href {\doibase 10.1016/0370-2693(83)90841-9} {\bibfield
  {journal} {\bibinfo  {journal} {Phys. Lett. B}\ }\textbf {\bibinfo {volume}
  {129}},\ \bibinfo {pages} {193} (\bibinfo {year}
  {1983}{\natexlab{b}})}\BibitemShut {NoStop}%
\bibitem [{\citenamefont {Hosotani}(1989)}]{Hosotani:1988bm}%
  \BibitemOpen
  \bibfield  {author} {\bibinfo {author} {\bibfnamefont {Y.}~\bibnamefont
  {Hosotani}},\ }\href {\doibase 10.1016/0003-4916(89)90015-8} {\bibfield
  {journal} {\bibinfo  {journal} {Annals Phys.}\ }\textbf {\bibinfo {volume}
  {190}},\ \bibinfo {pages} {233} (\bibinfo {year} {1989})}\BibitemShut
  {NoStop}%
\bibitem [{\citenamefont {Hatanaka}\ \emph {et~al.}(1998)\citenamefont
  {Hatanaka}, \citenamefont {Inami},\ and\ \citenamefont
  {Lim}}]{Hatanaka:1998yp}%
  \BibitemOpen
  \bibfield  {author} {\bibinfo {author} {\bibfnamefont {H.}~\bibnamefont
  {Hatanaka}}, \bibinfo {author} {\bibfnamefont {T.}~\bibnamefont {Inami}}, \
  and\ \bibinfo {author} {\bibfnamefont {C.}~\bibnamefont {Lim}},\ }\href
  {\doibase 10.1142/S021773239800276X} {\bibfield  {journal} {\bibinfo
  {journal} {Mod. Phys. Lett. A}\ }\textbf {\bibinfo {volume} {13}},\ \bibinfo
  {pages} {2601} (\bibinfo {year} {1998})},\ \Eprint
  {http://arxiv.org/abs/hep-th/9805067} {arXiv:hep-th/9805067} \BibitemShut
  {NoStop}%
\bibitem [{\citenamefont {Haba}\ \emph {et~al.}(2008)\citenamefont {Haba},
  \citenamefont {Matsumoto}, \citenamefont {Okada},\ and\ \citenamefont
  {Yamashita}}]{Haba:2008dr}%
  \BibitemOpen
  \bibfield  {author} {\bibinfo {author} {\bibfnamefont {N.}~\bibnamefont
  {Haba}}, \bibinfo {author} {\bibfnamefont {S.}~\bibnamefont {Matsumoto}},
  \bibinfo {author} {\bibfnamefont {N.}~\bibnamefont {Okada}}, \ and\ \bibinfo
  {author} {\bibfnamefont {T.}~\bibnamefont {Yamashita}},\ }\href {\doibase
  10.1143/PTP.120.77} {\bibfield  {journal} {\bibinfo  {journal} {Prog. Theor.
  Phys.}\ }\textbf {\bibinfo {volume} {120}},\ \bibinfo {pages} {77} (\bibinfo
  {year} {2008})},\ \Eprint {http://arxiv.org/abs/0802.3431} {arXiv:0802.3431
  [hep-ph]} \BibitemShut {NoStop}%
\bibitem [{\citenamefont {Maru}\ and\ \citenamefont
  {Yamashita}(2006)}]{Maru:2006wa}%
  \BibitemOpen
  \bibfield  {author} {\bibinfo {author} {\bibfnamefont {N.}~\bibnamefont
  {Maru}}\ and\ \bibinfo {author} {\bibfnamefont {T.}~\bibnamefont
  {Yamashita}},\ }\href {\doibase 10.1016/j.nuclphysb.2006.07.023} {\bibfield
  {journal} {\bibinfo  {journal} {Nucl. Phys. B}\ }\textbf {\bibinfo {volume}
  {754}},\ \bibinfo {pages} {127} (\bibinfo {year} {2006})},\ \Eprint
  {http://arxiv.org/abs/hep-ph/0603237} {arXiv:hep-ph/0603237} \BibitemShut
  {NoStop}%
\bibitem [{\citenamefont {Hosotani}\ \emph {et~al.}(2007)\citenamefont
  {Hosotani}, \citenamefont {Maru}, \citenamefont {Takenaga},\ and\
  \citenamefont {Yamashita}}]{Hosotani:2007kn}%
  \BibitemOpen
  \bibfield  {author} {\bibinfo {author} {\bibfnamefont {Y.}~\bibnamefont
  {Hosotani}}, \bibinfo {author} {\bibfnamefont {N.}~\bibnamefont {Maru}},
  \bibinfo {author} {\bibfnamefont {K.}~\bibnamefont {Takenaga}}, \ and\
  \bibinfo {author} {\bibfnamefont {T.}~\bibnamefont {Yamashita}},\ }\href
  {\doibase 10.1143/PTP.118.1053} {\bibfield  {journal} {\bibinfo  {journal}
  {Prog. Theor. Phys.}\ }\textbf {\bibinfo {volume} {118}},\ \bibinfo {pages}
  {1053} (\bibinfo {year} {2007})},\ \Eprint {http://arxiv.org/abs/0709.2844}
  {arXiv:0709.2844 [hep-ph]} \BibitemShut {NoStop}%
\bibitem [{\citenamefont {Hisano}\ \emph {et~al.}(2020)\citenamefont {Hisano},
  \citenamefont {Shoji},\ and\ \citenamefont {Yamada}}]{Hisano:2019cxm}%
  \BibitemOpen
  \bibfield  {author} {\bibinfo {author} {\bibfnamefont {J.}~\bibnamefont
  {Hisano}}, \bibinfo {author} {\bibfnamefont {Y.}~\bibnamefont {Shoji}}, \
  and\ \bibinfo {author} {\bibfnamefont {A.}~\bibnamefont {Yamada}},\ }\href
  {\doibase 10.1007/JHEP02(2020)193} {\bibfield  {journal} {\bibinfo  {journal}
  {JHEP}\ }\textbf {\bibinfo {volume} {02}},\ \bibinfo {pages} {193} (\bibinfo
  {year} {2020})},\ \Eprint {http://arxiv.org/abs/1908.09158} {arXiv:1908.09158
  [hep-ph]} \BibitemShut {NoStop}%
\bibitem [{\citenamefont {Scrucca}\ \emph {et~al.}(2003)\citenamefont
  {Scrucca}, \citenamefont {Serone},\ and\ \citenamefont
  {Silvestrini}}]{Scrucca:2003ra}%
  \BibitemOpen
  \bibfield  {author} {\bibinfo {author} {\bibfnamefont {C.~A.}\ \bibnamefont
  {Scrucca}}, \bibinfo {author} {\bibfnamefont {M.}~\bibnamefont {Serone}}, \
  and\ \bibinfo {author} {\bibfnamefont {L.}~\bibnamefont {Silvestrini}},\
  }\href {\doibase 10.1016/j.nuclphysb.2003.07.013} {\bibfield  {journal}
  {\bibinfo  {journal} {Nucl. Phys. B}\ }\textbf {\bibinfo {volume} {669}},\
  \bibinfo {pages} {128} (\bibinfo {year} {2003})},\ \Eprint
  {http://arxiv.org/abs/hep-ph/0304220} {arXiv:hep-ph/0304220} \BibitemShut
  {NoStop}%
\bibitem [{\citenamefont {Cacciapaglia}\ \emph {et~al.}(2006)\citenamefont
  {Cacciapaglia}, \citenamefont {Csaki},\ and\ \citenamefont
  {Park}}]{Cacciapaglia:2005da}%
  \BibitemOpen
  \bibfield  {author} {\bibinfo {author} {\bibfnamefont {G.}~\bibnamefont
  {Cacciapaglia}}, \bibinfo {author} {\bibfnamefont {C.}~\bibnamefont {Csaki}},
  \ and\ \bibinfo {author} {\bibfnamefont {S.~C.}\ \bibnamefont {Park}},\
  }\href {\doibase 10.1088/1126-6708/2006/03/099} {\bibfield  {journal}
  {\bibinfo  {journal} {JHEP}\ }\textbf {\bibinfo {volume} {03}},\ \bibinfo
  {pages} {099} (\bibinfo {year} {2006})},\ \Eprint
  {http://arxiv.org/abs/hep-ph/0510366} {arXiv:hep-ph/0510366} \BibitemShut
  {NoStop}%
\bibitem [{\citenamefont {Adachi}\ and\ \citenamefont
  {Maru}(2018)}]{Adachi:2018gud}%
  \BibitemOpen
  \bibfield  {author} {\bibinfo {author} {\bibfnamefont {Y.}~\bibnamefont
  {Adachi}}\ and\ \bibinfo {author} {\bibfnamefont {N.}~\bibnamefont {Maru}},\
  }\href@noop {} {\  (\bibinfo {year} {2018})},\ \Eprint
  {http://arxiv.org/abs/1809.02748} {arXiv:1809.02748 [hep-ph]} \BibitemShut
  {NoStop}%
\bibitem [{\citenamefont {Funatsu}\ \emph {et~al.}(2020)\citenamefont
  {Funatsu}, \citenamefont {Hatanaka}, \citenamefont {Hosotani}, \citenamefont
  {Orikasa},\ and\ \citenamefont {Yamatsu}}]{Funatsu:2020znj}%
  \BibitemOpen
  \bibfield  {author} {\bibinfo {author} {\bibfnamefont {S.}~\bibnamefont
  {Funatsu}}, \bibinfo {author} {\bibfnamefont {H.}~\bibnamefont {Hatanaka}},
  \bibinfo {author} {\bibfnamefont {Y.}~\bibnamefont {Hosotani}}, \bibinfo
  {author} {\bibfnamefont {Y.}~\bibnamefont {Orikasa}}, \ and\ \bibinfo
  {author} {\bibfnamefont {N.}~\bibnamefont {Yamatsu}},\ }\href {\doibase
  10.1103/PhysRevD.102.015005} {\bibfield  {journal} {\bibinfo  {journal}
  {Phys. Rev. D}\ }\textbf {\bibinfo {volume} {102}},\ \bibinfo {pages}
  {015005} (\bibinfo {year} {2020})},\ \Eprint
  {http://arxiv.org/abs/2002.09262} {arXiv:2002.09262 [hep-ph]} \BibitemShut
  {NoStop}%
\bibitem [{\citenamefont {Panico}\ \emph {et~al.}(2006)\citenamefont {Panico},
  \citenamefont {Serone},\ and\ \citenamefont {Wulzer}}]{Panico:2005dh}%
  \BibitemOpen
  \bibfield  {author} {\bibinfo {author} {\bibfnamefont {G.}~\bibnamefont
  {Panico}}, \bibinfo {author} {\bibfnamefont {M.}~\bibnamefont {Serone}}, \
  and\ \bibinfo {author} {\bibfnamefont {A.}~\bibnamefont {Wulzer}},\ }\href
  {\doibase 10.1016/j.nuclphysb.2006.01.025} {\bibfield  {journal} {\bibinfo
  {journal} {Nucl. Phys. B}\ }\textbf {\bibinfo {volume} {739}},\ \bibinfo
  {pages} {186} (\bibinfo {year} {2006})},\ \Eprint
  {http://arxiv.org/abs/hep-ph/0510373} {arXiv:hep-ph/0510373} \BibitemShut
  {NoStop}%
\bibitem [{\citenamefont {Fujii}\ \emph {et~al.}(2019)\citenamefont {Fujii}
  \emph {et~al.}}]{Fujii:2019zll}%
  \BibitemOpen
  \bibfield  {author} {\bibinfo {author} {\bibfnamefont {K.}~\bibnamefont
  {Fujii}} \emph {et~al.} (\bibinfo {collaboration} {LCC Physics Working
  Group}),\ }\href@noop {} {\  (\bibinfo {year} {2019})},\ \Eprint
  {http://arxiv.org/abs/1908.11299} {arXiv:1908.11299 [hep-ex]} \BibitemShut
  {NoStop}%
\bibitem [{\citenamefont {Roloff}\ \emph {et~al.}(2020)\citenamefont {Roloff},
  \citenamefont {Schnoor}, \citenamefont {Simoniello},\ and\ \citenamefont
  {Xu}}]{Roloff:2019crr}%
  \BibitemOpen
  \bibfield  {author} {\bibinfo {author} {\bibfnamefont {P.}~\bibnamefont
  {Roloff}}, \bibinfo {author} {\bibfnamefont {U.}~\bibnamefont {Schnoor}},
  \bibinfo {author} {\bibfnamefont {R.}~\bibnamefont {Simoniello}}, \ and\
  \bibinfo {author} {\bibfnamefont {B.}~\bibnamefont {Xu}} (\bibinfo
  {collaboration} {CLICdp}),\ }\href {\doibase 10.1140/epjc/s10052-020-08567-7}
  {\bibfield  {journal} {\bibinfo  {journal} {Eur. Phys. J. C}\ }\textbf
  {\bibinfo {volume} {80}},\ \bibinfo {pages} {1010} (\bibinfo {year}
  {2020})},\ \Eprint {http://arxiv.org/abs/1901.05897} {arXiv:1901.05897
  [hep-ex]} \BibitemShut {NoStop}%
\bibitem [{\citenamefont {Antoniadis}\ \emph {et~al.}(2001)\citenamefont
  {Antoniadis}, \citenamefont {Benakli},\ and\ \citenamefont
  {Quiros}}]{Antoniadis:2001cv}%
  \BibitemOpen
  \bibfield  {author} {\bibinfo {author} {\bibfnamefont {I.}~\bibnamefont
  {Antoniadis}}, \bibinfo {author} {\bibfnamefont {K.}~\bibnamefont {Benakli}},
  \ and\ \bibinfo {author} {\bibfnamefont {M.}~\bibnamefont {Quiros}},\ }\href
  {\doibase 10.1088/1367-2630/3/1/320} {\bibfield  {journal} {\bibinfo
  {journal} {New J. Phys.}\ }\textbf {\bibinfo {volume} {3}},\ \bibinfo {pages}
  {20} (\bibinfo {year} {2001})},\ \Eprint
  {http://arxiv.org/abs/hep-th/0108005} {arXiv:hep-th/0108005} \BibitemShut
  {NoStop}%
\bibitem [{\citenamefont {Panico}\ \emph {et~al.}(2007)\citenamefont {Panico},
  \citenamefont {Serone},\ and\ \citenamefont {Wulzer}}]{Panico:2006em}%
  \BibitemOpen
  \bibfield  {author} {\bibinfo {author} {\bibfnamefont {G.}~\bibnamefont
  {Panico}}, \bibinfo {author} {\bibfnamefont {M.}~\bibnamefont {Serone}}, \
  and\ \bibinfo {author} {\bibfnamefont {A.}~\bibnamefont {Wulzer}},\ }\href
  {\doibase 10.1016/j.nuclphysb.2006.10.032} {\bibfield  {journal} {\bibinfo
  {journal} {Nucl. Phys. B}\ }\textbf {\bibinfo {volume} {762}},\ \bibinfo
  {pages} {189} (\bibinfo {year} {2007})},\ \Eprint
  {http://arxiv.org/abs/hep-ph/0605292} {arXiv:hep-ph/0605292} \BibitemShut
  {NoStop}%
\bibitem [{\citenamefont {Cepeda}\ \emph {et~al.}(2019)\citenamefont {Cepeda}
  \emph {et~al.}}]{Cepeda:2019klc}%
  \BibitemOpen
  \bibfield  {author} {\bibinfo {author} {\bibfnamefont {M.}~\bibnamefont
  {Cepeda}} \emph {et~al.},\ }\href {\doibase 10.23731/CYRM-2019-007.221}
  {\bibfield  {journal} {\bibinfo  {journal} {CERN Yellow Rep. Monogr.}\
  }\textbf {\bibinfo {volume} {7}},\ \bibinfo {pages} {221} (\bibinfo {year}
  {2019})},\ \Eprint {http://arxiv.org/abs/1902.00134} {arXiv:1902.00134
  [hep-ph]} \BibitemShut {NoStop}%
\end{thebibliography}%

\end{document}